\documentclass[conference]{IEEEtran}
\usepackage[ruled,vlined,linesnumbered]{algorithm2e}
\usepackage{xcolor}
\usepackage{booktabs}
\usepackage{multirow}
\usepackage[utf8]{inputenc}
\usepackage{amsmath, amssymb}
\usepackage{graphicx}
\usepackage{cite}
\usepackage{geometry}
\usepackage[mathscr]{euscript}
\usepackage{multirow}
\begin{document}

\title{Multi-Antenna Channel Prediction in 5G and Beyond using Data driven approach }
\author{\IEEEauthorblockN{A. Sathi Babu}
\IEEEauthorblockA{\textit{Department of ECE } \\
\textit{SRM University AP}\\
Guntur, India \\
sathibabu\_arepalli@srmap.edu.in }\\
\and
\IEEEauthorblockN{V. Udaya Sankar }
\IEEEauthorblockA{\textit{Department of ECE } \\
\textit{SRM University AP}\\
Guntur, India \\
udayasankar.v@srmap.edu.in }\\
\and
\IEEEauthorblockN{Vishnu Ram OV}
\IEEEauthorblockA{\textit{Independent Research Consultant} \\
Bengaluru, India \\
vishnu.n@ieee.org}
}
\date{}
\maketitle
\begin{abstract}
As Wireless Communication evolves towards 5G and beyond, it satisfies Quality of Service (QoS) requirements of users by utilizing beam forming, adaptive transmission and multi antenna communication techniques. It is essential to have accurate channel information at Base station (BS) to provide satisfactory QoS to users. Traditional channel estimation approach involves transmitting pilots from User Equipments (UEs) and estimating the channel at BS, increases computational complexity as it is periodic process. It also increases communication complexity as number of antennas at BS increases which leads to increase in pilots. In this paper, we consider channel vector prediction using Machine learning approach at 7GHz. In this work, we generated data using ray tracing mechanism and used this generated data to train Machine learning models. We Considered XG Boost regression and Multilayer perceptron regression model.  
\end{abstract}. 


\section{Introduction}

Accurate channel information at Transmitter (Tx) or Receiver (Rx) is essential in 5G and beyond wireless systems to satisfy Quality of service (QoS) of users. High datarate, low latency, low bit error rate (BER), reliable communication are some of QoS metrics that can be satisfied using adaptive transmission, beamforming and
multi-antenna communication techniques. In conventional multiple antenna wireless systems, channel estimation is performed using pilot symbols and estimate complexity increases with increase in number of antennas\cite{massivemimobook}. In addition, increasing the number of antennas leads to an increase in pilot overhead and a decrease in data throughput. This motivates us to look at channel estimation methods using data driven approaches. In \cite{li2025survey}, the authors studied channel estimation methods using a traditional and data driven approach. 

In \cite{kim2020massive}, the authors addressed problems related to channel prediction in massive multiple input and multiple output (MIMO) systems by developing and comparing multilayer perceptron (MLP) and vector based Kalman filtering. The emerging $7$--$24$ GHz upper-mid-band (known as FR3) has increasing attention due to medium coverage and capacity enhancement along with it offers opportunities for MIMO beamforming\cite{cui20256g}. 

A deterministic ray tracing (RT) that characterizes the radio propagation channel is studied in \cite{7152831} and an RT based dataset is generated to use for Machine Learning (ML) model to predict channel characteristics is provided in \cite{alkhateeb2019deepmimo}. The Sionna RT provides a GPU-accelerated RT framework for radio propagation modeling that enables channel responses to be generated by considering Tx-Rx locations, antenna configurations, antenna characteristics and the surrounding propagation environment \cite{hoydis2023sionna}. A survey of environmental aware channel knowledge map for 6G is provided in \cite{zeng2024tutorial}. 

In \cite{xu2026mixture}, authors propose a novel framework that consists of mixture-of-experts for channel prediction that uses an hybrid dataset (incorporates both Line of sight and Non Line of sight) at carrier frequency of 2.4GHz. A multimodal flow-matching framework for pilot-free channel inference that integrates camera images, LiDAR point clouds, and GPS coordinates to estimate channel state information (CSI) in dynamic wireless environments is studied in \cite{liang2026pilot}. Whereas in \cite{tumma2025deep}, authors studied a pilot free channel estimation using variational autoencoders in Unmanned Aerial Vehicle-Free Space Optical systems environment. An ML (Random Forest, K-nearest neighbour) based path loss prediction using an RT generated synthetic datasets is provided in  \cite{tarhouni2025machine}. In this paper, we consider baseband complex channel vector at Base station (BS) using data driven approach for a given User Equipment (UE) location. A data set is generated using an RT model and to predict channel coefficient vector we used an ML approach. 

The organization of the paper is as follows. Section II introduces the system model, followed by solution approach in Section III, the Section IV describes simulation results and Section V concludes the paper. 
\section{System Model}

\begin{figure}[htbp]
    \centering   \includegraphics[width=0.9\linewidth]{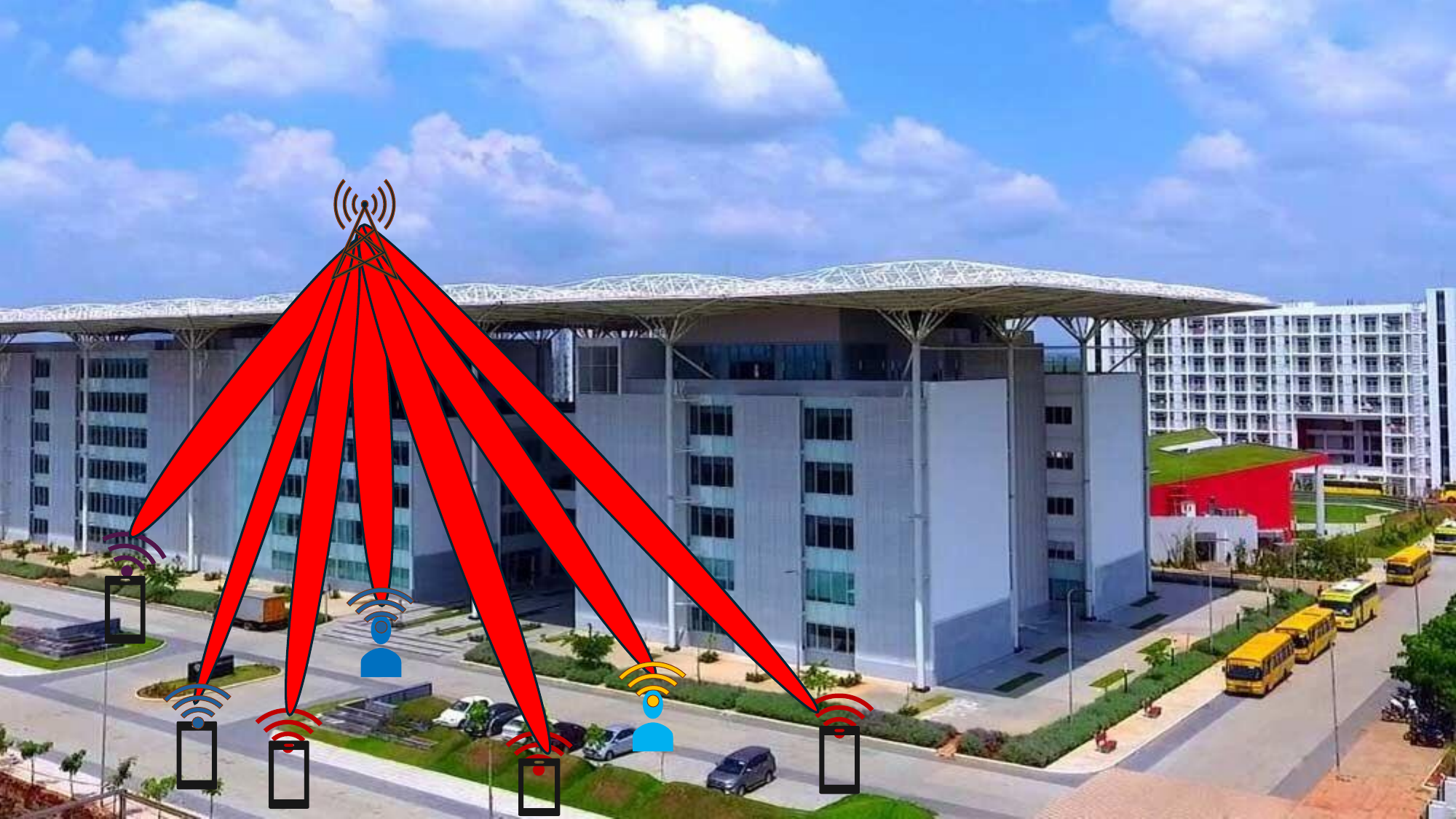}
    \caption{ A rooftop SBS $N_t\times1$ MISO setup with multiple UEs}
\label{SRM_OUTDOOR}
\end{figure}

We consider a small base station (SBS), mounted on the rooftop of a building that serves multiple users with single antena user equipment (UE) (Fig \ref{SRM_OUTDOOR}). The SBS has $N_T$-antennas used for beamforming in Downlink (DL). We consider a flat fading (static) channel between each SBS antenna and UE. We consider a Line of Sight (LoS) points from each SBS antenna but has LOS and Non-LOS transmission paths. Also, we consider a single carrier transmission between each SBS antenna (Transmitter-Tx) and UE (Receiver-Rx).

Let $\mathbf{h}^i = \left[h^i_1, h^i_2, ..., h^i_{N_T} \right]^T$ be the transmit antenna array from SBS to $i^{th}$ UE, where $h^i_m = h^i_{m,re} + jh^i_{m,im}, m=1,2,...,N_T$ is a baseband complex channel coefficient of $m^{th}$ transmit antenna of SBS. Let $x^i$ be the complex baseband transmit symbol from SBS to $i^{th}$ UE, then  
\begin{equation}
y_i=\mathbf{h}^{{i}^{H}}\mathbf{w}x^i + n_i
\label{eq:received_signal}
\end{equation}
where, $n_i \sim \mathcal{CN}(0,\sigma_n^2)$ and $\mathbf{w} \in C^{N_T \times 1}$ is a complex transmit beamforming weight vector. It can be computed using Maximal Ratio Transmission (MRT)/Matched Filtering (MF) or Zero Forcing (ZF) beamforming which is a function of $\mathbf{h}^i$ (i.e., $f(\mathbf{h}^i)$). Traditionally, $\mathbf{h}^i$ is estimated by UE using pilot symbols or SBS can estimate using pilots sent by UE. This estimation process is continuous \& periodic which leads to increase in communication complexity and computational complexity as $N_T$ increases. 

An advantage of replacing traditional blocks in wireless communication with AI enabled blocks is studied in \cite{hoydis2021toward}. The major important block in wireless communication system is channel estimation. In this work we consider transmit channel vector prediction using data-driven approach. 

\section{Solution Approach}

Our objective is to predict $N_T$ channels in $N_T\times1$ MISO framework for a given set of UE locations at SBS. We assume UE location is available at the SBS.  The process consists of generating channel coefficient vector ($\mathbf{h}^i$) and use these data to train the Machine Learning (ML) model that predicts channel vector for a given UE location. 

\subsection{$N_T\times1$ MISO Data (Channel Coefficient) Generation}

We consider a Sionna  ray tracing (RT) engine to generate vector of narrowband ccomplex channel coefficients that corresponds to $N_T$ SBS transmit antenna elements. A summary of data-generation procedure is given Algorithm~1. The coordinate posistion of SBS (Tx) and UEs (Rx) are first converted into UTM coordinates and subsequently mapped to local cartesian coordinates using Blender/Sionna RT since Sionan RT works only with local coordinates. For each pair of SBS and $i^{th}$ UE location, the vector of $\mathbf{h}^i \in \mathcal{C}^{N_T\times 1}$ MISO channel coefficient is generated using Sionna raytracing engine and the resulting complex channel coefficients are stored along with Tx, Rx location.  

\begin{algorithm}[t]
\caption{$N_T\times1$ MISO Dataset Generation}
\label{alg:miso_dataset}
\footnotesize
\SetKwInOut{Input}{Input}
\SetKwInOut{Output}{Output}

\Input{Outdoor scene, Tx position $\mathbf{P}_{Tx}$,
Rx positions $\{\mathbf{P}^i_{Rx}\}_{i=1}^{L}$, carrier frequency $f_c$}

\Output{$\mathscr{D}
=\{(\mathbf{P}_{Tx},\mathbf{P}^i_{Rx},\mathbf{h}^i_{re},\mathbf{h}^i_{im})\}_{i=1}^{L}$}

Model the outdoor environment using outdoor scene\;

Convert geographical Tx--Rx coordinates to UTM coordinates\;

Map UTM coordinates to local Cartesian coordinates (using Blender/Sionna)\;

Configure Sionna RT with a $N_T\times1$ MISO antenna setup\;

\For{$i=1$ \KwTo $L$}{
    Place the UE at $i^{th}$ position $\mathbf{P}^{Rx}_{i}$ and execute ray tracing\;
    \For{$m=1$ \KwTo $N_T$}{
        Extract multipath parameters
        $(\alpha^{i,m}_n,\tau^{i,m}_n)$ for $n \in \mathscr{N}^i_m$, where $\mathscr{N}^i_m$ is set of all multipath components for $m^{th}$ SBS antenna to $i^{th}$ UE location\;

        Compute the narrowband channel coefficient $\textbf{h}^{m,i}$ using eq. \ref{eq:miso_channel}\;
    }
    Store
    $(\mathbf{P}_{Tx},\mathbf{P}^i_{Rx},
    \mathbf{h}^i_{re},\mathbf{h}^i_{im}) \in \mathscr{D}$, where $\mathbf{h}^i_{re} = [h^i_{m,re}]_{m=1}^{N_T}$ and $\mathbf{h}^i_{im} = [h^i_{m,im}]_{m=1}^{N_T}$\;
}

\Return $\mathscr{D}$\;

\end{algorithm}

Let $\alpha^{i,m}_n$ be the $n^{th}$ path gain of $i^{th}$ UE location from SBS antenna $m$ and $\tau^{i,m}_n$ denotes the $n^{th}$ path propagation delay of $i^{th}$ UE location from SBS antenna $m$. Then, the multipath parameters $(\alpha^{i,m}_n,\tau^{i,m}_n)$ are generated
using Sionna RT and corresponding channel impulse response $h^{i,m}(\tau)$ for $i^{th}$ UE location is expressed as

\begin{equation}
h^i_m(\tau)
=
\sum_{n \in \mathscr{N}^i_m}
\alpha^{i,m}_n\,
\delta\!\left(\tau-\tau^{i,m}_n\right),
m\in\{1,2,...,N_T\}.
\label{eq:miso_cir}
\end{equation}

For a given carrier frequency $f_c$, the narrow band complex channel coefficient between the $m$-th SBS antenna and the $i^{th}$ UE location is given by 

\begin{equation}
h^i_m
=
\sum_{n\in\mathscr{N}^i_m}
\alpha^{i,m}_n
e^{-j2\pi f_c \tau^{i,m}_n},
\qquad m\in\{1,2,...,N_T\},
\label{eq:miso_channel}
\end{equation}

The resulting dataset contains the local transmitter coordinates
$\mathbf{P}_{{Tx}}$, receiver coordinates
$\mathbf{P}^i_{Rx}$, corresponding real and imaginary components of 
MISO channel vector. The generated dataset using Sionna RT is given as 

\begin{equation}
\mathscr{D}
=
\left\{
\left(
\mathbf{P}_{Tx},
\mathbf{P}^i_{Rx},
\mathbf{h}^i_{re},
\mathbf{h}^i_{im}
\right)
\right\}_{i=1}^{L} \sim iid 
\quad \mathscr{P}_{\bar{\mathbf{X}},\bar{\mathbf{h}}}, 
\label{eq:miso_dataset}
\end{equation} 
Where, $\mathbf{P}_{Tx} = (P_{x,Tx},P_{y,Tx},P_{z,Tx})$, $\mathbf{P}^i_{Rx} = (P^i_{x,Rx},P^i_{y,Rx},P^i_{z,Rx})$, $\bar{\mathbf{X}}^i =(\mathbf{P}_{Tx},\mathbf{P}_{Rx}^i)$ is $i^{th}$ feature and $\bar{\mathbf{h}}^i=\left(\mathbf{h}^i_{re},\mathbf{h}^i_{im}\right)$ is target vector. The data set ($\bar{\mathbf{X}}^i,\bar{\mathbf{h}}^i)$ for $i=1,2,...L$ forms data set with multiple target variable and is used to train multi target regression model. 

\subsection{Machine-Learning-Based Channel Prediction}
\label{subsec:ml_channel_prediction}

A study of development of machine learning algorithms via probabilistic approach is given in \cite{murphy2022probabilistic} and also channel prediciton strategies using regression approch is studied in \cite{electronics12234740}. 

We consider $\mathscr{D}_{re} =\left\{\left(\mathbf{P}_{Tx},\mathbf{P}^i_{Rx},\mathbf{h}^i_{re} \right)\right\}_{i=1}^L \sim \mathscr{P}_{\mathbf{\bar{X}},\mathbf{\bar{h}}_{re}}$ and $\mathscr{D}_{im} =\left\{\left(\mathbf{P}_{Tx},\mathbf{P}^i_{Rx},\mathbf{h}^i_{im} \right)\right\}_{i=1}^L \sim \mathscr{P}_{\mathbf{\bar{X}},\mathbf{\bar{h}}_{im}}$ to train ML models for regression separately as real and imaginary part of channel vectors are independent but not have an identical distribution. Lets consider dataset $\mathscr{D}_{re}$ and our objective is to 
predict $P_{\frac{\mathbf{\bar{h}_{re}}}{\mathbf{\bar{X}}}}$ from $\mathscr{D}_{re}$. 
An ML based approach is used to predict channel as described below.

\begin{figure}[htbp]
\centering
\includegraphics[width=0.8\columnwidth]{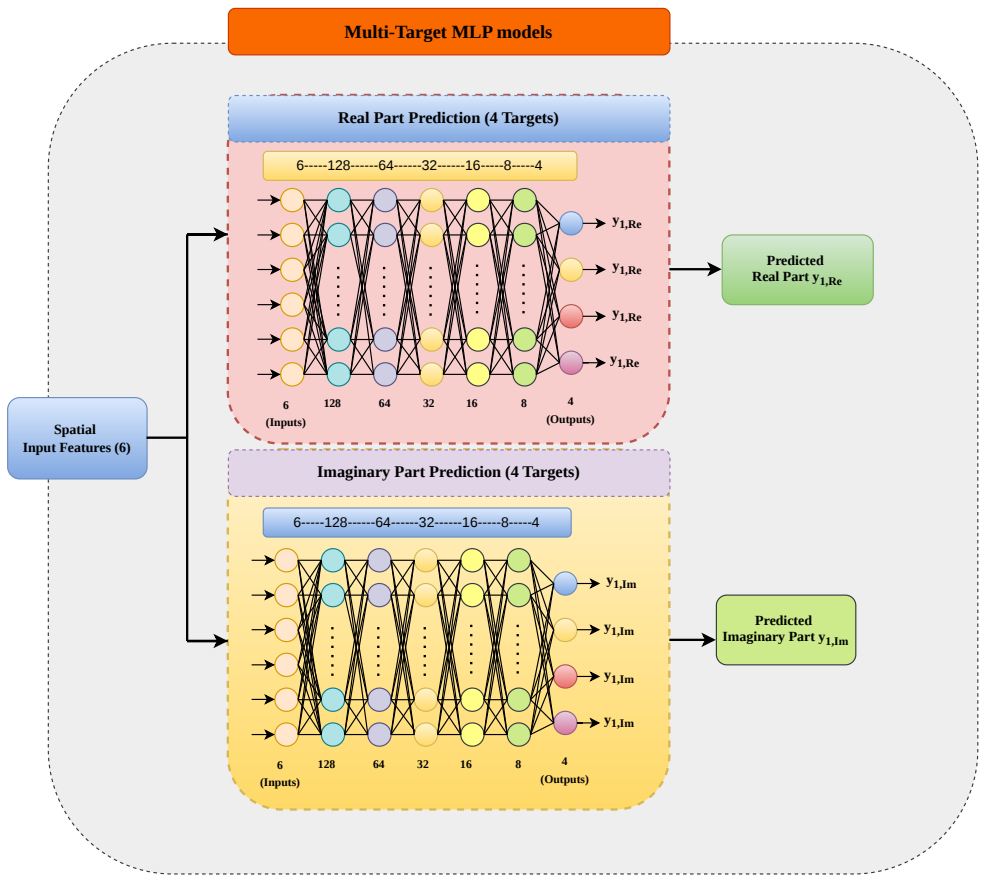}
\caption{An MLP architecture}
\label{fig:MLP_FINAL}
\end{figure}

The dataset $\mathscr{D}_{re}$ is randomly divided into 80\% training and 20\% testing samples using a random seed. We consider XG-Boost regression and Multilayer perceptron (MLP) for regression since data may be non-linearly separable as Support vector regression (SVR) with RBF kernel, K-Nearest Neighbor (KNN) regression, Random Forest gave very poor results when we tested. As XG-Boost is tree based model, we did not use any standardization on data set where as for MLP we used Min-Max scalar for the dataset  $\mathscr{D}_{re}$. We consider MLP model for $N_T = 4$ as shown in Fig. \ref{fig:MLP_FINAL}. It consists of $5$ hiddenlayers of sizes $128,64,32,16,8$ with input layer of size $6$ that consider $\bar{\mathbf{X}}^i$ as input and $\mathbf{\bar{h}_{re}}$ as output $\mathbf{\bar{y}}_{re}$. We considered ReLu activation function for hidden layers and linear for output layer. 

The prediction performance of the model is evaluated on unseen test samples using root
mean square error (RMSE), mean absolute error (MAE), coefficient of
determination ($R^2$), and adjusted coefficient of determination
($R_{\mathrm{Adj}}^2$). Let $n$ be the number of test samples, let $y_i$ be target variable, predicted value be $\hat{y}_i$, and $k$ be number of input features, then 

\begin{equation}
\mathrm{RMSE}
=
\sqrt{
\frac{1}{n}
\sum_{i=1}^{n}
\left(y_i-\hat{y}_i\right)^2
},
\label{eq:rmse}
\end{equation}

\begin{equation}
\mathrm{MAE}
=
\frac{1}{n}
\sum_{i=1}^{n}
\left|y_i-\hat{y}_i\right|,
\label{eq:mae}
\end{equation}

\begin{equation}
R^2
=
1-
\frac{
\sum_{i=1}^{n}
\left(y_i-\hat{y}_i\right)^2
}{
\sum_{i=1}^{n}
\left(y_i-\bar{y}\right)^2
},
\label{eq:r2}
\end{equation}

and

\begin{equation}
R_{\mathrm{Adj}}^{2}
=
1-
\frac{(1-R^2)(n-1)}
{n-k-1}.
\label{eq:adjusted_r2}
\end{equation}

Here, proportion of variance in test data is given by $R^2$ metric and adjusted $R^2$ accounts for number of input features. We considered a large carrier carrier frequency $f_c (\approx 7GHz)$ to generate data $\mathscr{D}=\left\{\left(\mathbf{P}_{Tx},\mathbf{P}^i_{Rx},
\mathbf{h}^i_{re},\mathbf{h}^i_{im}\right)\right\}_{i=1}^{L}$. A distance of $\frac{\lambda}{2} (\approx 2.14cm)$ causes change in $\pi$-radian phase shift in the corresponding channel coefficient $\mathbf{\bar{h}}$. Hence, a small change is UE location for a fixed SBS location leads to large variations in $\mathbf{h}_{re}$ and $\mathbf{h}_{im}$ value which makes model fitting very poor (which we mentioned in results). We apply phase compensation to the data before model training and testing as given below.  

Let $\mathbf{h}^i = \mathbf{h}^i_{re} + j\mathbf{h}^i_{im}$, then $\mathbf{g}^i = \mathbf{g}^i_{re} + j\mathbf{g}^i_{im} = \mathbf{h}^ie^{\frac{j2\pi f_cd_i}{c}}$. Where Euclidian distance betwen SBS and $i^{th}$ UE location is give by $d_i = ||\mathbf{P}_{Tx}-\mathbf{P}^i_{Rx}||_2$. The resulting dataset becomes $\mathscr{D}_{re} =\left\{\left(\mathbf{P}_{Tx},\mathbf{P}^i_{Rx},\mathbf{g}^i_{re} \right)\right\}_{i=1}^L$ which can be used for model traininga nd testing. It is easy to obtain $\hat{\mathbf{h}}^i_{re}$ from $\hat{\mathbf{h}}^i_{re}$ using $\hat{\mathbf{h}}^i_{re}=\hat{\mathbf{g}}^i_{re} e^{\frac{-j2\pi f_cd_i}{c}}$. This process adjusts the dominant distance-dependent carrier phase while preserving relative amplitude and phase relationships among all channel coefficients. Similar approach we used for another dataset  $\mathscr{D}_{im} =\left\{\left(\mathbf{P}_{Tx},\mathbf{P}^i_{Rx},\mathbf{h}^i_{im} \right)\right\}_{i=1}^L$. 
\section{Simulation Results}

The proposed framework combines physically consistent 3D environment modelling in Blender with the NVIDIA Sionna RT engine for wireless channel data generation. We considered a $0.3 \times 0.3$~km outdoor university campus environment (latitude and longitude coordinate ranges: 16.46269—16.46564, 80.50635-–80.50887) with realistic building geometry, roads, vegetation, terrain and material properties suitable for propagation at 7~GHz. The assigned material and surface properties enable characterization of geometry-dependent multipath effects. We consider the following simulation parameters for our experiments as given in Table \ref{tab:simulation_parameters_outdoor}. 


\begin{table}[h]
\caption{Simulation Configuration for Outdoor Scenario}
\label{tab:simulation_parameters_outdoor}
\centering
\begin{tabular}{lc}
\hline
\textbf{Parameter} & \textbf{Value} \\
\hline
Carrier Frequency & 7\,GHz \\

Environment Size & $0.3 \times 0.3$ km \\

Transmitter Height & 21.44 m \\

Receiver Height & 1.5 m \\

Antenna Type & Isotropic \\

Transmitter Antenna Array & Planar Array with $N_T=4$ \\

Receiver Antenna Array & Planar Array $(1 \times 1)$ \\

Antenna Polarization & Vertical \\

Random Seed & 42\\

\hline
\end{tabular}
\end{table}

\begin{figure}[htbp]
    \centering
    \includegraphics[width=0.9\linewidth]{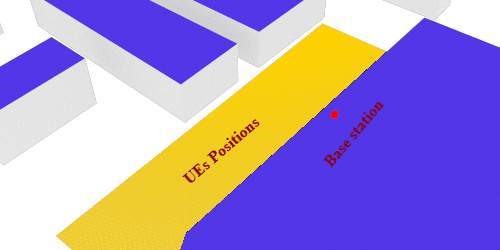}
    \caption{The campus with a fixed Tx position and 10,000 Rx locations}
    \label{fig:srm_outdoor_model}
\end{figure}

We considered approximately $10000$ receiver positions across the proposed region to generate data points using Sionna Ray-tracing framework \cite{aoudia2025sionnart} and Sionna RT simulates multipath propagation between SBS antenna to receiver locations. The Open-StreetMap is used to import building footprints. The receiver height is considered as 1.5\.m according to the 3GPP channel modeling \cite{3gpp38901R17}. Also, we considered a uniformly distributed UE positions across the proopsed region for a given fixed SBS transmitter (Fig. \ref{fig:srm_outdoor_model}) to capture spatial channel variations.

\begin{figure}[h]
\centering
\includegraphics[width=\linewidth]{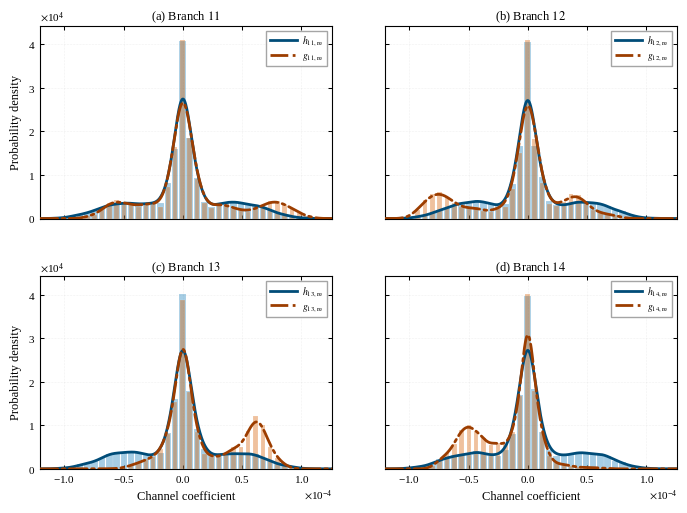}
\caption{PDF distributions of $\mathbf{h}_{re},\mathbf{g}_{re}$}
\label{fig:pdf_h_re_and_g_re}
\end{figure}

The probability density function (PDF) of real component of channel coefficient $\mathbf{h}_{re}$ and phase compensated channel coefficient $\mathbf{g}_{re}$ is shown in Fig. ~\ref{fig:pdf_h_re_and_g_re}.It is observed that distribution for $\mathbf{h}_{re}$ is almost normal where as for $\mathbf{g}_{re}$ is multi-model for each components due to phase compensation. The corresponding statistical properties are given in Tables \ref{tab:h_statistics}, \ref{tab:g_statistics}. We can observe that variance for $\mathbf{h}_{re}$ is almost similar with different mean values, whereas for $\mathbf{g}_{re}$ we can see that slight change in the variance along with different mean. 

\begin{table}[htbp]
\centering
\caption{Statistical Properties of $\mathbf{h}_{re},\mathbf{h}_{im}$}
\label{tab:h_statistics}

\footnotesize

\begin{tabular}{lcc}
\hline
\textbf{Component} & \textbf{Mean} & \textbf{Variance} \\
\hline

$h_{1,re}$ & $-2.335494 \times 10^{-7}$ & $1.166674 \times 10^{-9}$ \\
$h_{1,im}$ & $ 3.166148 \times 10^{-7}$ & $1.186250 \times 10^{-9}$ \\

$h_{2,re}$ & $ 3.171154 \times 10^{-7}$ & $1.177761 \times 10^{-9}$ \\
$h_{2,im}$ & $-2.445272 \times 10^{-7}$ & $1.163628 \times 10^{-9}$ \\

$h_{3,re}$ & $-5.624366 \times 10^{-7}$ & $1.158319 \times 10^{-9}$ \\
$h_{3,im}$ & $-9.911796 \times 10^{-8}$ & $1.166690 \times 10^{-9}$ \\

$h_{4,re}$ & $ 2.114224 \times 10^{-7}$ & $1.164199 \times 10^{-9}$ \\
$h_{4,im}$ & $ 4.661824 \times 10^{-7}$ & $1.146148 \times 10^{-9}$ \\

\hline
\end{tabular}

\end{table}

\begin{table}[htbp]
\centering
\caption{Statistical Properties of $\mathbf{g}_{re},\mathbf{g}_{im}$}
\label{tab:g_statistics}

\footnotesize

\begin{tabular}{lcc}
\hline
\textbf{Component} & \textbf{Mean} & \textbf{Variance} \\
\hline

$g_{1,re}$ & $ 5.677706 \times 10^{-6}$ & $1.383203 \times 10^{-9}$ \\
$g_{1,im}$ & $-5.498493 \times 10^{-6}$ & $9.074002 \times 10^{-10}$ \\

$g_{2,re}$ & $-9.067630 \times 10^{-6}$ & $1.406073 \times 10^{-9}$ \\
$g_{2,im}$ & $ 7.153780 \times 10^{-6}$ & $8.020642 \times 10^{-10}$ \\

$g_{3,re}$ & $ 1.868802 \times 10^{-5}$ & $9.660787 \times 10^{-10}$ \\
$g_{3,im}$ & $-7.473835 \times 10^{-7}$ & $1.009421 \times 10^{-9}$ \\

$g_{4,re}$ & $-1.655669 \times 10^{-5}$ & $6.776940 \times 10^{-10}$ \\
$g_{4,im}$ & $-1.748513 \times 10^{-5}$ & $1.053003 \times 10^{-9}$ \\

\hline
\end{tabular}

\end{table}

\begin{figure}[h]
\centering
\includegraphics[width=\linewidth]{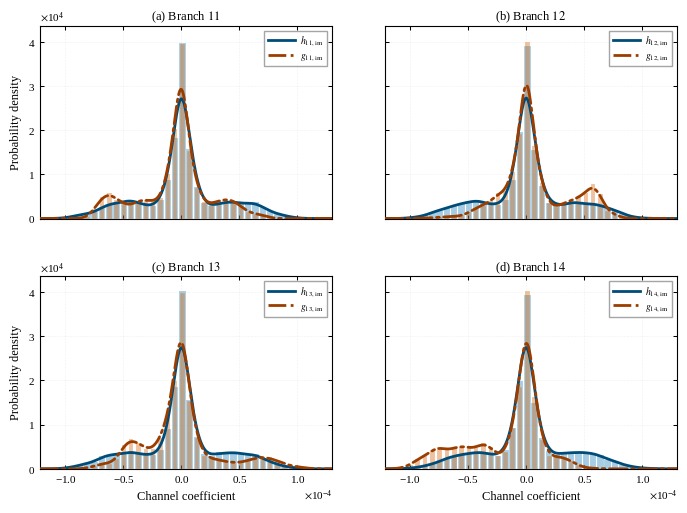}
\caption{PDF distributions of $\mathbf{h}_{im},\mathbf{g}_{im}$}
\label{fig:pdf_h_im_and_g_im}
\end{figure}

We consider a randomly selected 8000-data samples from 10000 generated dataset for model training and remaining 2000 samples for model testing to test generalization capability of the model. We consider the following model training parameters for MLP as given in Table \ref{tab:model_training_mlp}. 

\begin{table}[h]
\caption{Model training parameters for MLP}
\label{tab:model_training_mlp}
\centering
\begin{tabular}{lc}
\hline
\textbf{Parameter} & \textbf{Value} \\
\hline
Loss function & SmoothL1 (Huber) Loss \\

Huber parameter ($\beta$) & 0.5 \\

Data split & Random 80:20 \\

Optimizer & AdamW \\

Learning rate ($\eta$) & 0.001 \\

Batch Size & 128 \\

Random seed & 42 \\

Epochs & 500\\

\hline
\end{tabular}
\end{table}

\begin{figure}[h]
\centering
\includegraphics[width=\columnwidth]{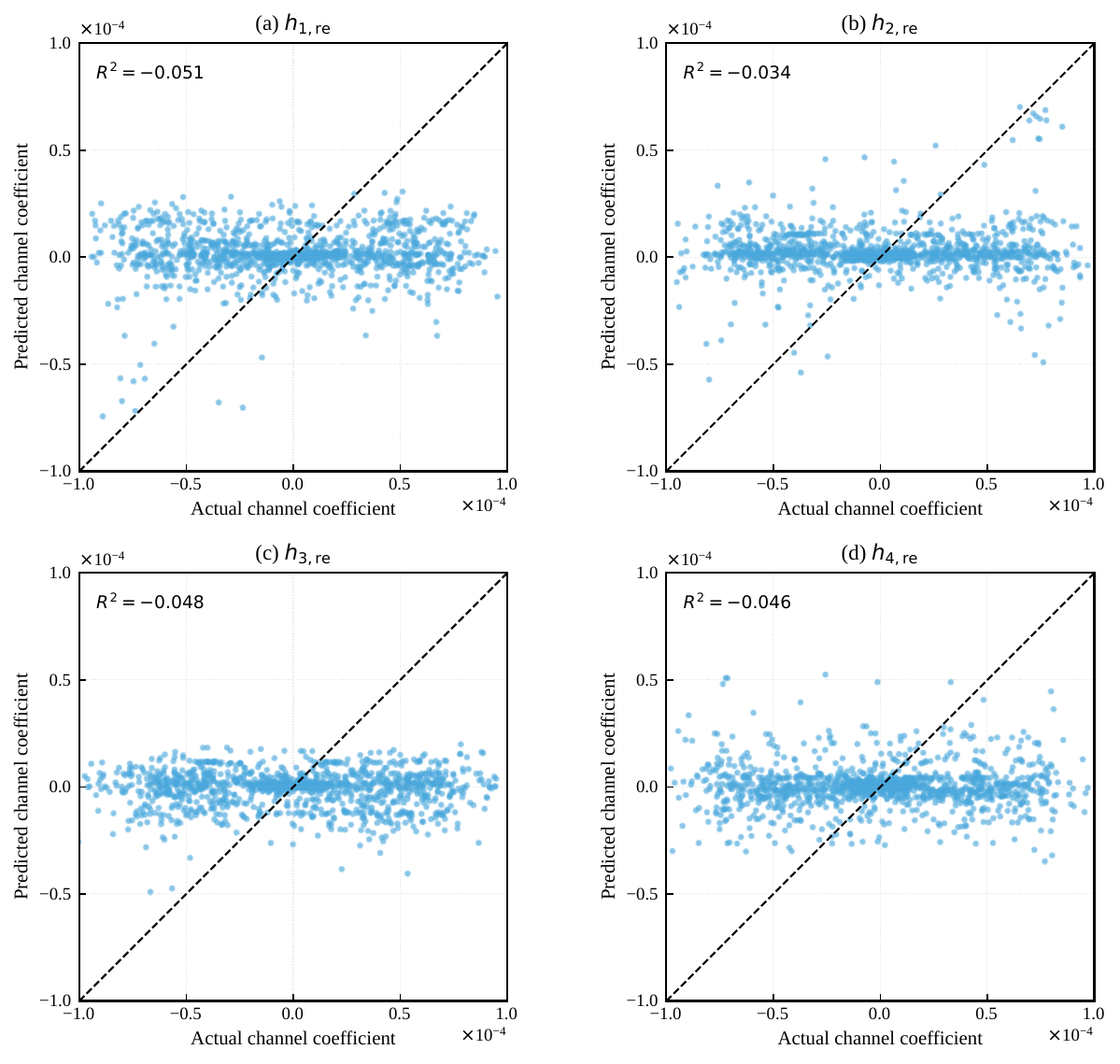}
\caption{
Actual versus predicted channel coefficients using  MLP for
(a) $h_{1,re}$ (b) $h_{2,re}$ (c) $h_{3,re}$ (d) $h_{4,re}$
}
\label{fig:direct_random_real}
\end{figure}

\begin{figure}[h]
\centering
\includegraphics[width=\columnwidth]{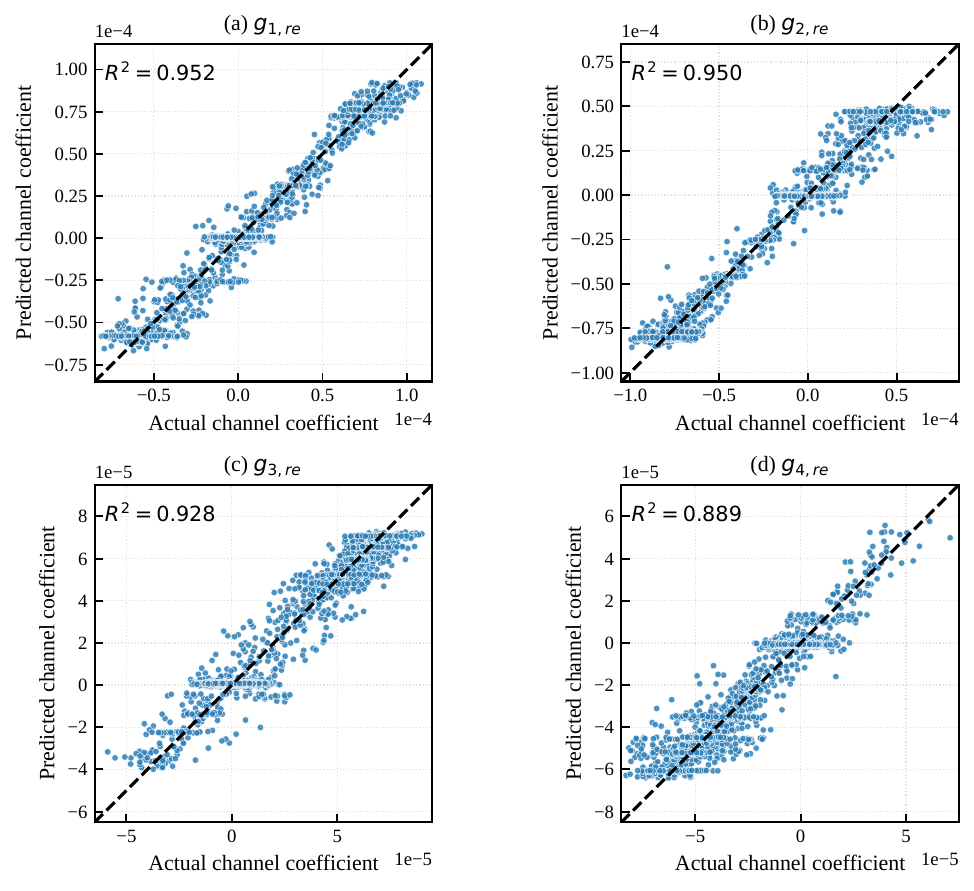}
\caption{
Actual versus predicted channel coefficients using MLP for
(a) $g_{1,re}$ (b) $g_{2,re}$ (c) $g_{3,re}$ (d) $g_{4,re}$
}
\label{fig:phase_random_real}
\end{figure}

\begin{figure}[h]
\centering
\includegraphics[width=\columnwidth]{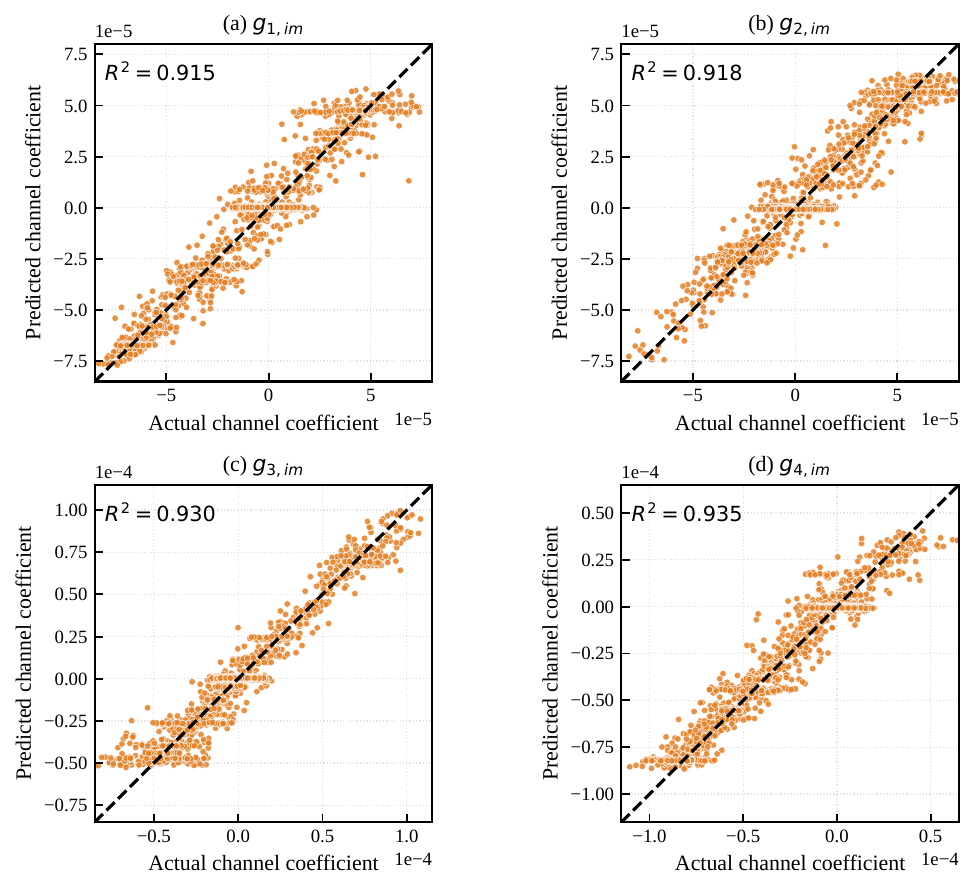}
\caption{
Actual versus predicted channel coefficients using MLP for 
(a) $g_{1,im}$ (b) $g_{2,im}$ (c) $g_{3,im}$ (d) $g_{4,im}$.
}
\label{fig:phase_random_imaginary}
\end{figure}

\begin{figure}[h]
\centering
\includegraphics[width=\columnwidth]{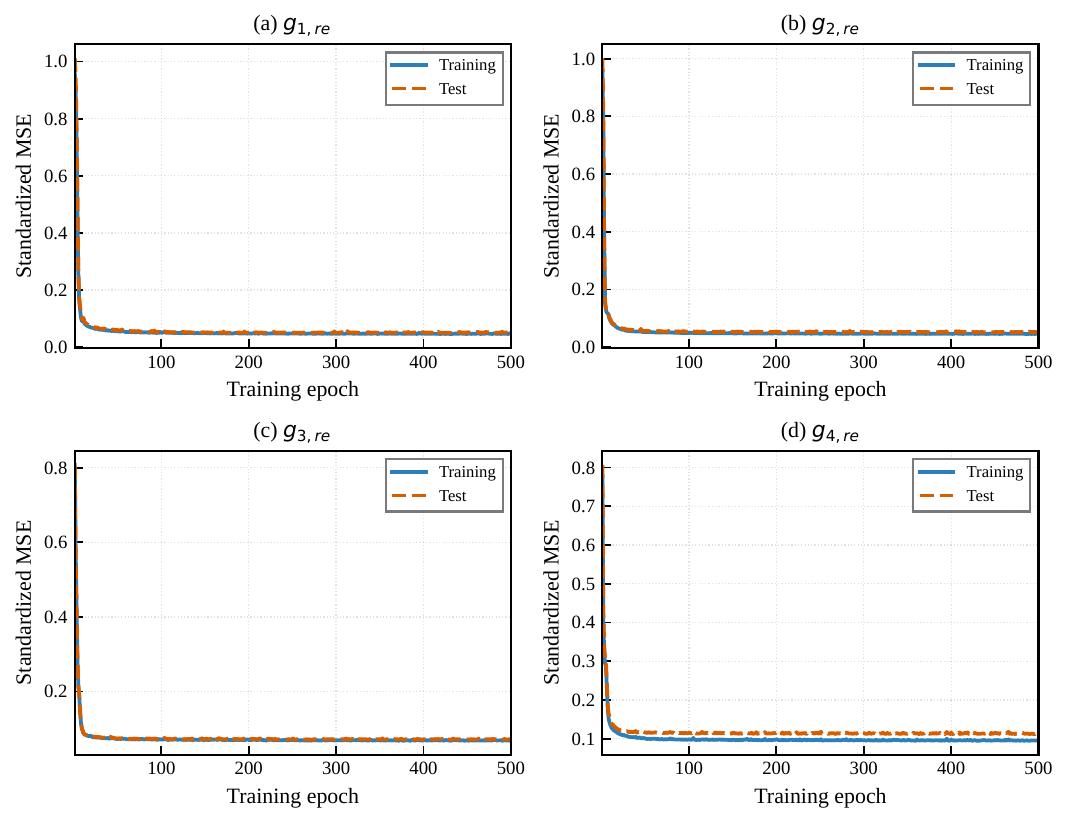}
\caption{
Training and Test SmoothL1 loss convergence of MLP for
(a) $g_{1,re}$ (b) $g_{2,re}$ (c) $g_{3,re}$ (d) $g_{4,re}$
}
\label{fig:g_re_losscurve_mlp_randomsplit}
\end{figure}

\begin{figure}[h]
\centering
\includegraphics[width=\columnwidth]{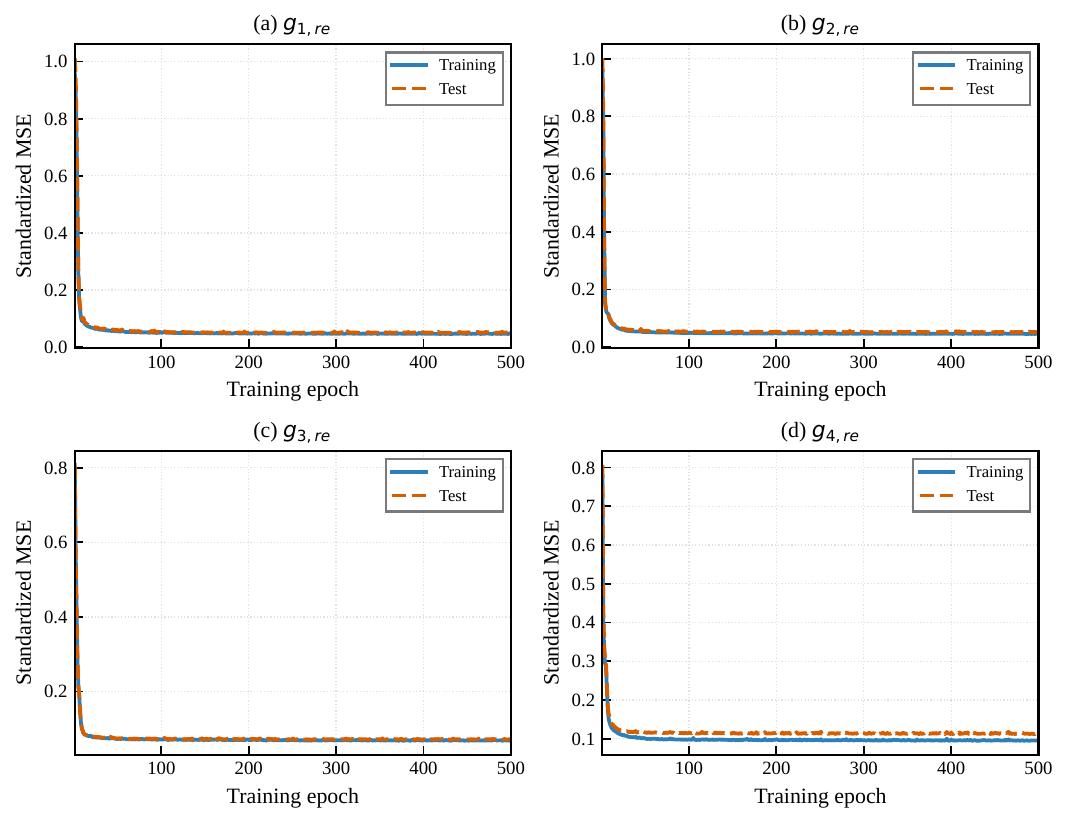}
\caption{
Training and Test SmoothL1 loss convergence of MLP for 
(a) $g_{1,im}$ (b) $g_{2,im}$ (c) $g_{3,im}$ (d) $g_{4,im}$
}
\label{fig:g_im_losscurve_mlp_randomsplit}
\end{figure}

The Table~\ref{tab:direct_h_branch_performance} summarizes the performance of the trained model for $\mathbf{h}_{re}$. It is observed that $R^2$ value is negative the test data that signifies poor data fitting. The Fig~\ref{fig:direct_random_real} shows relation between actual and predicted values which is not diagonal that describes poor data fitting \cite{murphy2022probabilistic}. Similarly, performance of the trained models for $\mathbf{g}_{re}$ is given in the Table~\ref{tab:phase_branch_performance}. It is observed that MLP constantly outperforms XGBoost with $R^2$ values of 0.9507, 0.9482, 0.9237, 0.8887 for $g_{1,re}, g_{2,re}, g_{3,re}, g_{4,re}$ respectively and corresponding RMSE values are $8.449\times10^{-6}$, $8.655\times10^{-6}$, $8.593\times10^{-6}$, $8.766\times10^{-6}$. We can see from Fig ~ \ref{fig:phase_random_real} that relation between actual and predicted samples are diagonal which specifies better data fit for the MLP model \cite{murphy2022probabilistic}. Also, Fig ~\ref{fig:g_re_losscurve_mlp_randomsplit} shows convergence cures of training and testing error for proposed MLP. Similar result we can observe for $\mathbf{h}_{im}$ and $\mathbf{g}_{im}$ as shown in Figs ~\ref{fig:phase_random_imaginary}, \ref{fig:g_im_losscurve_mlp_randomsplit}. 

\begin{table*}[!t]
\centering
\caption{
Performance of the MLP for $\mathbf{h}_{re},\mathbf{h}_{im}$
}
\label{tab:direct_h_branch_performance}

\footnotesize
\renewcommand{\arraystretch}{1.08}
\setlength{\tabcolsep}{4pt}

\begin{tabular*}{\textwidth}
{@{\extracolsep{\fill}}cclcccc@{}}

\toprule

\textbf{Target} &
\textbf{Branch} &
\textbf{Model} &
$\mathbf{R^2}$ &
\textbf{Adjusted }$\mathbf{R^2}$ &
\textbf{MAE} &
\textbf{RMSE} \\

\midrule

\multirow{4}{*}{$h_{\mathrm{real}}$}

& $h_{1,re}$
& MLP
& $-0.057443$
& $-0.060627$
& $2.409\times10^{-5}$
& $3.563\times10^{-5}$ \\

\cmidrule(lr){2-7}

& $h_{2,re}$
& MLP
& $-0.034012$
& $-0.037125$
& $2.478\times10^{-5}$
& $3.617\times10^{-5}$ \\

\cmidrule(lr){2-7}

& $h_{3,re}$
& MLP
& $-0.047913$
& $-0.051068$
& $2.496\times10^{-5}$
& $3.613\times10^{-5}$ \\

\cmidrule(lr){2-7}

& $h_{4,re}$
& MLP
& $-0.045962$
& $-0.049111$
& $2.379\times10^{-5}$
& $3.524\times10^{-5}$ \\

\midrule

\multirow{4}{*}{$h_{\mathrm{imag}}$}

& $h_{1,im}$
& MLP
& $-0.028260$
& $-0.031355$
& $2.429\times10^{-5}$
& $3.541\times10^{-5}$ \\

\cmidrule(lr){2-7}

& $h_{2,im}$
& MLP
& $-0.059832$
& $-0.063022$
& $2.375\times10^{-5}$
& $3.521\times10^{-5}$ \\

\cmidrule(lr){2-7}

& $h_{3,im}$
& MLP
& $-0.036258$
& $-0.039378$
& $2.388\times10^{-5}$
& $3.481\times10^{-5}$ \\

\cmidrule(lr){2-7}

& $h_{4,im}$
& MLP
& $-0.075664$
& $-0.078903$
& $2.450\times10^{-5}$
& $3.609\times10^{-5}$ \\

\bottomrule

\end{tabular*}
\end{table*}

\begin{table*}[!t]
\centering
\caption{
Performance of the XGBoost and MLP for $\mathbf{g}_{re},\mathbf{g}_{im}$
}
\label{tab:phase_branch_performance}

\footnotesize
\renewcommand{\arraystretch}{1.08}
\setlength{\tabcolsep}{4pt}

\begin{tabular*}{\textwidth}
{@{\extracolsep{\fill}}cclcccc@{}}

\toprule

\textbf{Target} &
\textbf{Branch} &
\textbf{Model} &
$\mathbf{R^2}$ &
\textbf{Adjusted }$\mathbf{R^2}$ &
\textbf{MAE} &
\textbf{RMSE} \\

\midrule

\multirow{8}{*}{$g_{\mathrm{re}}$}

& \multirow{2}{*}{$g_{1,re}$}
& XGBoost
& 0.931961
& 0.931756
& $6.875\times10^{-6}$
& $9.923\times10^{-6}$ \\

&
& \textbf{MLP}
& \textbf{0.951633}
& \textbf{0.951488}
& \textbf{$6.222\times10^{-6}$}
& \textbf{$8.366\times10^{-6}$} \\

\cmidrule(lr){2-7}

& \multirow{2}{*}{$g_{2,re}$}
& XGBoost
& 0.932240
& 0.932036
& $6.863\times10^{-6}$
& $9.894\times10^{-6}$ \\

&
& \textbf{MLP}
& \textbf{0.950211}
& \textbf{0.950061}
& \textbf{$6.253\times10^{-6}$}
& \textbf{$8.481\times10^{-6}$} \\

\cmidrule(lr){2-7}

& \multirow{2}{*}{$g_{3,re}$}
& XGBoost
& 0.894922
& 0.894606
& $6.980\times10^{-6}$
& $1.009\times10^{-5}$ \\

&
& \textbf{MLP}
& \textbf{0.928301}
& \textbf{0.928086}
& \textbf{$6.162\times10^{-6}$}
& \textbf{$8.331\times10^{-6}$} \\

\cmidrule(lr){2-7}

& \multirow{2}{*}{$g_{4,re}$}
& XGBoost
& 0.838157
& 0.837670
& $7.333\times10^{-6}$
& $1.057\times10^{-5}$ \\

&
& \textbf{MLP}
& \textbf{0.889073}
& \textbf{0.888739}
& \textbf{$6.444\times10^{-6}$}
& \textbf{$8.752\times10^{-6}$} \\

\midrule

\multirow{8}{*}{$g_{\mathrm{im}}$}

& \multirow{2}{*}{$g_{1,im}$}
& XGBoost
& 0.878142
& 0.877775
& $7.310\times10^{-6}$
& $1.050\times10^{-5}$ \\

&
& \textbf{MLP}
& \textbf{0.914739}
& \textbf{0.914483}
& \textbf{$6.465\times10^{-6}$}
& \textbf{$8.785\times10^{-6}$} \\

\cmidrule(lr){2-7}

& \multirow{2}{*}{$g_{2,im}$}
& XGBoost
& 0.876364
& 0.875992
& $7.090\times10^{-6}$
& $1.019\times10^{-5}$ \\

&
& \textbf{MLP}
& \textbf{0.918029}
& \textbf{0.917782}
& \textbf{$6.187\times10^{-6}$}
& \textbf{$8.294\times10^{-6}$} \\

\cmidrule(lr){2-7}

& \multirow{2}{*}{$g_{3,im}$}
& XGBoost
& 0.899733
& 0.899432
& $7.125\times10^{-6}$
& $1.032\times10^{-5}$ \\

&
& \textbf{MLP}
& \textbf{0.929663}
& \textbf{0.929451}
& \textbf{$6.292\times10^{-6}$}
& \textbf{$8.642\times10^{-6}$} \\

\cmidrule(lr){2-7}

& \multirow{2}{*}{$g_{4,im}$}
& XGBoost
& 0.910674
& 0.910405
& $6.712\times10^{-6}$
& $9.801\times10^{-6}$ \\

&
& \textbf{MLP}
& \textbf{0.935112}
& \textbf{0.934916}
& \textbf{$6.168\times10^{-6}$}
& \textbf{$8.354\times10^{-6}$} \\

\bottomrule

\end{tabular*}

\end{table*}

\section{Conclusion}

In this work, we presented multi-channel prediction framework for $N_T=4$ at 7GHz frequency band. We generated dataset consists of channel vector as multi-target variable for a given set of transmitter and receiver locations. The generated dataset is used to train the XGBoost and MLP regression models. It is observed from the Table~\ref{tab:phase_branch_performance}, that both models fits data very well with $>90\%$ of $R^2$ value except for $g_{4,re}$ channel parameter. It is also observe that the MLP outperforms in comparison with that of the XGBoost regression. We have to consider a tradeoff between explainability (XGBoost) and performance (MLP) in choosing the model depending on a scenario. Looking forward, we focus on multi model approach that considers image of a region as an additional input to improve performance of the model $>97\%$ for channel vector prediction including $g_{4,re}$ and also test for multiple use case scenarios.   
\section*{Acknowledgment}
The authors would like to thank the International Telecommunication Union (ITU) AI for Good for their valuable meetings and discussions.

\bibliographystyle{IEEEtran}
\bibliography{references}

\end{document}